\documentclass[12pt]{article}
\usepackage{amsmath}
\usepackage{graphicx}
\usepackage{bm}
\usepackage{fancyhdr}
\usepackage{amssymb}
\usepackage{caption}
\usepackage{subfigure}
\begin{document}



\def\a{\alpha}
\def\b{\beta}
\def\d{\delta}
\def\e{\epsilon}
\def\g{\gamma}
\def\h{\mathfrak{h}}
\def\k{\kappa}
\def\l{\lambda}
\def\o{\omega}
\def\p{\wp}
\def\r{\rho}
\def\t{\tau}
\def\s{\sigma}
\def\z{\zeta}
\def\x{\xi}
\def\V={{{\bf\rm{V}}}}
 \def\A{{\cal{A}}}
 \def\B{{\cal{B}}}
 \def\C{{\cal{C}}}
 \def\D{{\cal{D}}}
\def\K{{\cal{K}}}
\def\O{\Omega}
\def\R{\bar{R}}
\def\T{{\cal{T}}}
\def\L{\Lambda}
\def\f{E_{\tau,\eta}(sl_2)}
\def\E{E_{\tau,\eta}(sl_n)}
\def\Zb{\mathbb{Z}}
\def\Cb{\mathbb{C}}

\def\R{\overline{R}}

\def\beq{\begin{equation}}
\def\eeq{\end{equation}}
\def\bea{\begin{eqnarray}}
\def\eea{\end{eqnarray}}
\def\ba{\begin{array}}
\def\ea{\end{array}}
\def\no{\nonumber}
\def\le{\langle}
\def\re{\rangle}
\def\lt{\left}
\def\rt{\right}

\newtheorem{Theorem}{Theorem}
\newtheorem{Definition}{Definition}
\newtheorem{Proposition}{Proposition}
\newtheorem{Lemma}{Lemma}
\newtheorem{Corollary}{Corollary}
\newcommand{\proof}[1]{{\bf Proof. }
        #1\begin{flushright}$\Box$\end{flushright}}

\baselineskip=20pt

\newfont{\elevenmib}{cmmib10 scaled\magstep1}
\newcommand{\preprint}{
   \begin{flushleft}
   \end{flushleft}\vspace{-1.3cm}
   \begin{flushright}\normalsize
   \end{flushright}}
\newcommand{\Title}[1]{{\baselineskip=26pt
   \begin{center} \Large \bf #1 \\ \ \\ \end{center}}}
\newcommand{\Author}{\begin{center}
   \large \bf
Fakai Wen${}^{a,b,c}$,~Junpeng Cao${}^{c,d,e}\footnote{Corresponding author: junpengcao@iphy.ac.cn}$,~Tao Yang${}^{a,b}$,~ Kun Hao${}^{a,b}$, Zhan-Ying Yang${}^{b,f}$ and Wen-Li
Yang${}^{a,b}\footnote{Corresponding author: wlyang@nwu.edu.cn}$
 \end{center}}
\newcommand{\Address}{\begin{center}

     ${}^a$Institute of Modern Physics, Northwest University,
     Xi'an 710069, China\\
     ${}^b$Shaanxi Key Laboratory for Theoretical Physics Frontiers,
Xi'an 710069, China\\
     ${}^c$Institute of Physics, Chinese
Academy of Sciences, Beijing 100190, China\\
     ${}^d$School of Physical Sciences, University of Chinese Academy of
Sciences, Beijing, China\\
     ${}^e$Collaborative Innovation Center of Quantum Matter, Beijing,
     China\\
     ${}^f$School of Physics, Northwest University, Xi'an 710069, China\\

   \end{center}}

\preprint \thispagestyle{empty}
\bigskip\bigskip\bigskip

\Title{Surface energy of the one-dimensional supersymmetric $t-J$ model with unparallel boundary fields} \Author

\Address \vspace{1cm}

\begin{abstract}
We investigate the thermodynamic limit of the exact solution, which is given by an  inhomogeneous $T-Q$ relation, of the one-dimensional supersymmetric $t-J$ model with unparallel boundary magnetic fields. It is shown that the contribution of the inhomogeneous term at the ground state satisfies the $L^{-1}$ scaling law, where $L$ is the system-size.
This fact enables us to calculate the surface (or boundary) energy of the
system. The method used in this paper can be generalized to study the thermodynamic
limit and surface energy of other models related to rational R-matrices.

\vspace{1truecm} \noindent {\it PACS:} 75.10.Pq; 02.30.Ik; 71.10.Pm


\noindent {\it Keywords}: The supersymmetric $t-J$ model; Bethe ansatz; $T-Q$ relation; Thermodynamic limit;
Surface energy
\end{abstract}

\newpage



\section{Introduction}
\label{intro} \setcounter{equation}{0}
The $t-J$ model  is the strongly repulsive limit of the well-known Hubbard model \cite{Physica.86.375, J.Phys.C.Solid.State.Phys.10.L271,Phys.Rev.B.37.3759}, which has played a fundamental and important role in strongly correlated electronic systems. The model is also one of the cornerstone models in the study of high-$T_{c}$ superconductivity \cite{Science.235.1196, Phys.Rev.Lett.68.2960, PhysRevB.91.195132, PhysRevLett.116.067002}. In general, the  Hamiltonian includes nearest-neighbor
hopping ($t$) and nearest-neighbor spin exchange and charge interactions ($J$) (see below (\ref{Hamiltonian_basic})) for the periodic case \cite{PhysRevB.46.9147}.
For the open case, the Hamiltonian also includes the boundary chemical potentials $\chi_{1},\chi_{L}$ and  the boundary fields $\mathbf{h}_{1},\mathbf{h}_{L}$ \cite{PhysRevB.61.3450, Nucl.Phys.B.777.352}, i.e.,
\begin{eqnarray}
&&H=-t\sum_{\alpha,j=1}^{L-1}\mathcal{P}\left[c_{j,\alpha}^{\dag}c_{j+1,\alpha}+c_{j+1,\alpha}^{\dag}c_{j,\alpha}\right]\mathcal{P}
+J\sum_{j=1}^{L-1}\left[\mathbf{S}_{j}\cdot \mathbf{S}_{j+1}-\frac{1}{4}n_{j}n_{j+1}\right] \nonumber\\
&&\quad\quad+\chi_{1}n_1
+2\mathbf{h}_{1}\cdot \mathbf{S}_{1}
+\chi_{L}n_{L}
+2\mathbf{h}_{L}\cdot \mathbf{S}_{L}, \label{Hamiltonian_basic}
\end{eqnarray}
where $L$ is the total number of lattice sites and the coupling constants  $\chi_1,\,\chi_L$ and $\mathbf{h}_{1},\,\mathbf{h}_{L}$ are given by (\ref{parameter_correspondences}) below. The operators $c_{j,\alpha}$ and $c_{j,\alpha}^{\dag}$ are the annihilation and creation operators of the electron with spin $\alpha=\pm1$ on the lattice site $j$, which satisfies anticommutation relations, i.e., $\left\{c_{i,\alpha}^{\dag},c_{j,\tau}\right\}=\delta_{i,j}\delta_{\alpha,\tau}$.
There are only three possible states at the lattice site $i$ due to the factor $\mathcal{P}=(1-n_{j,-\alpha})$ projects out double occupancies. The operator $n_{j}=\sum_{\alpha=\pm1}n_{j,\alpha}$ means the total number operator on site $j$ and $n_{j,\alpha}=c_{j,\alpha}^{\dag}c_{j,\alpha}$, and the total number operator of electrons $\hat{N}=\sum_{j=1}^{L}n_{j}$.
The spin operators $S=\sum_{j=1}^{L}S_{j}$,\quad $S^{\dag}=\sum_{j=1}^{L}S_{j}^{\dag}$ and $S^{z}=\sum_{j=1}^{L}S_{j}^{z}$ with the local operators:
$S_{j}=c_{j,1}^{\dag}c_{j,-1}$, $S_{j}^{\dag}=c_{j,-1}^{\dag}c_{j,1}$, $ S_{j}^{z}=\frac{1}{2}(n_{j,1}-n_{j,-1})$  form the $su(2)$ algebra.

At the supersymmetric points $J=\pm 2 t$, the Hamiltonian in one spatial dimension is supersymmetric and integrable \cite{Skl88, Phys.Rev.Lett.60.821, Phys.Rev.Lett.63.2140, Phys.Rev.B.46.9234Nucl.Phys.B.396.611, Phys.Rev.B.12.3795, Phys.Rev.B.36.5177, Phys.Rev.Lett.64.2831}. One can obtain the exact solution of the one-dimensional supersymmetric $t-J$ model with periodic boundary condition or parallel boundary fields by the nested Bethe ansatz method \cite{PhysRevB.46.9147, PhysRevB.61.3450} or the off-shell Bethe ansatz \cite{Mod.Phys.Lett.A.9.2029, J.Phys.A:Math.Theor.41.275202J.Phys.A:Math.Theor.45.055207}. Based on the exact solution, the properties of the $t-J$ models, for example, surface energy, the elementary excitation, the correlation functions and the thermodynamics have attracted a great attention \cite{Phys.Rev.Lett.93.036402,J.Phys.A:Math.Gen.29.6183, PhysRevB.66.245102, CambridgeUniversityPress.9780511524332}.
Compared with the periodic case and parallel boundary fields case, the one-dimensional supersymmetric $t-J$ model with unparallel boundary fields is the most general integrable case. With the help of  the exact solution of the one-dimensional supersymmetric $t-J$ model with unparallel boundary fields \cite{jstat14.04.P04031, wang2015off, JHEP.07.051}, the thermodynamic limit and surface energy of the model is a fascinating question\cite{PhysRevLett.111.137201, JPhysA.47.032001, NuclPhysB.884.17}.

In this paper, our goals are to study the thermodynamic limit and boundary effects of the supersymmetric $t-J$ model with unparallel boundary fields. Based on former works \cite{NuclPhysB.915.119}, one can not direct employ the thermodynamic Bethe ansatz (TBA) method to approach the thermodynamic limit of $t-J$ model due to the inhomogeneous term in the $T$-$Q$ relation. Therefore, the first thing should be addressed is the contribution of the inhomogeneous term.
In this paper, we choose  the region of $\chi_{1}>1$ and $\chi_{L}<1$ as an example.
Through the analysis of the
finite-lattice systems, it is shown that the contribution of the
inhomogeneous term in the associated $T-Q$ relation to the ground state energy satisfies the scaling law $L^{-1}$, where $L$ is the system-size.
Based on this fact, by using the standard thermodynamic Bethe ansatz method and taking the limit of temperature tending to zero, we find that all the Bethe roots are real at the ground state in the region of $\chi_{1}>1$ and $0\leq\chi_{L}<1$.
While in region of $\chi_{1}>1$ and $\chi_{L}<0$, besides the real Bethe roots, there exists the boundary bound state and the boundary bound state should be stable.
Furthermore, the surface energy of the system is calculated.
Comparison of the surface energy from
the analytic expressions with that from the Hamiltonian by the
extrapolation method, we show that they coincide with each other
very well.

The plan of the paper is as follows. We briefly  review  the Bethe ansatz solutions of the one-dimensional supersymmetric $t-J$ model with unparallel boundary fields in Section 2.
In Section 3, we focus on the contribution of the inhomogeneous term to the ground state energy.
In Section 4, with the help of the Bethe ansatz solution for the finite-size system,
we study the thermodynamic limit and surface energy of the model. We summarize our results and give
some discussions in Section 5.

\section{Bethe ansatz solutions}
\label{Bethe} \setcounter{equation}{0}

In this paper we consider $J=2t$ and $t=-1$, which corresponds to the supersymmetric and integrable point
\cite{PhysRevB.46.9147}. Let $\rm\bf V^{(m|n)}=\rm\bf V^{m}\oplus \rm\bf V^{n}$ denotes a graded linear space with
an orthonormal basis $\{|i\rangle,i=1,\cdots,m+n\}$ having the Grassmann parity (denoted by $\epsilon_i$):  $\epsilon_i=0$ for $i=1,\cdots,m$ and $\epsilon_i=1$ for $i=m+1,\cdots,m+n$,  which endows the fundamental representation of $su(m|n)$ algebra \cite{RevModPhys.47.573}.
For the supersymmetric $t-J$ model, we have $m=1$ and $n=2$ \cite{PhysRevB.46.9147}.
 The integrability of the model is associated with the $R$-matrix
\begin{eqnarray}
R_{i,j}(u)=u+\eta\Pi_{i,j},
\end{eqnarray}
where $u$ is the spectral parameter and $\eta$ is the crossing parameter, and $\Pi_{i,j}$ is the $Z_2$-graded permutation operator
\begin{eqnarray}
\left(\Pi_{i,j}\right)_{a_ia_j}^{b_ib_j}=\delta_{a_ib_j}\delta_{a_jb_i}(-1)^{\epsilon_{b_i}\epsilon_{b_j}}.
\end{eqnarray}
The $R$-matrix satisfies the graded Yang-Baxter equation
\begin{eqnarray}
 R_{12}(u-v)R_{13}(u)R_{23}(v)=R_{23}(v)R_{13}(u)R_{12}(u-v),
\end{eqnarray}
and possesses the properties:
\begin{eqnarray}
\mbox{Initial condition:}&&R_{12}(0)=\eta \Pi_{12}, \\
\mbox{Unitarity relation:}&&R_{12}(u)R_{21}(-u) = \rho_1(u)\,\times {\rm id},\\
\mbox{Crossing Unitarity relation:}&&R_{12}^{st_1}(-u+\eta)\,R_{21}^{st_1}(u)=\rho_2(u)\,\times {\rm id},
\end{eqnarray}
where $\rho_1(u)=-({u}-\eta)({u}+\eta)$, $\rho_2(u)=-{u}({u}-\eta)$, $R_{21}(u)=\Pi_{12}\,R_{21}(u)\,\Pi_{12}$ and $st_i$ denotes the super transposition in the $i$-th space $(A^{st})_{ij}=A_{ji}(-1)^{\epsilon_i[\epsilon_i+\epsilon_j]}$.
Here and below we adopt the standard notation: for any
matrix $A\in {\rm End}({\rm\bf V^{(m|n)}})$, $A_j$ is an super embedding operator
in the graded tensor product space ${\rm\bf V^{(m|n)}}\otimes {\rm\bf V^{(m|n)}}\otimes\cdots$,
which acts as $A$ on the $j$-th space and as an identity on the
other factor spaces; $R_{ij}(u)$ is an super embedding operator of
$R$-matrix in the graded tensor product space, which acts as an identity on the
factor spaces except for the $i$-th and $j$-th ones.

In this paper we consider the most general reflection matrices \footnote{Without losing the generalization, the $K^{\pm}(u)$ given by (\ref{Kmatrix}) and (\ref{dualKmatrix}) are the most
general $K$-matrices of the model and satisfy  $[K^{-}(u),K^+(v)]\neq 0$. This fact gives rise to  that  they cannot be diagonalized simultaneously (which corresponds to the non-diagonal (or unparallel) boundary fields), and that there does not exist  an obvious reference state on which the conventional Bethe ansatz \cite{wang2015off} can be performed.}:
\begin{eqnarray}
 K^{-}(u)=\left(
        \begin{array}{ccc}
          \xi+u & 0 & 0 \\
          0 & \xi+cu & 2c_1u \\
          0 & 2c_2u & \xi-cu \\
        \end{array}
      \right), \label{Kmatrix}
\end{eqnarray}
which satisfy the  reflection equation (RE)
\begin{eqnarray}
&&R_{12}(u-v)K_{1}^{-}(u)R_{21}(u+v)K_2^{-}(v)=K_2^{-}(v)R_{12}(u+v)K_{1}^{-}(u) R_{21}(u-v),\label{RE}
\end{eqnarray}
and
\begin{eqnarray}
 K^{+}(u)=\left(
        \begin{array}{ccc}
          \xi'-u & 0 & 0 \\
          0 & \xi'-\frac{\eta}{2}+c'\left(-u+\frac{\eta}{2}\right) & 2c_1'\left(-u+\frac{\eta}{2}\right) \\
          0 & 2c_2'\left(-u+\frac{\eta}{2}\right) & \xi'-\frac{\eta}{2}-c'\left(-u+\frac{\eta}{2}\right) \\
        \end{array}
      \right), \label{dualKmatrix}
\end{eqnarray}
which satisfies the dual RE respectively
\begin{eqnarray}
&&R_{12}(u-v)K_{1}^{+}(v)R_{21}(\eta-u-v)K_2^{+}(u) \nonumber\\
&&\qquad \qquad\qquad\qquad=K_2^{+}(u)R_{12}(\eta-u-v)K_{1}^{+}(v) R_{21}(u-v).\label{DRE}
\end{eqnarray}
The above parameters in (\ref{Kmatrix}) and (\ref{dualKmatrix}) have to satisfy the restrictions \cite{JHEP.07.051}
\begin{eqnarray}
c^2+4c_1c_2-1=0,~~~c'^2+4c_1'c_2'-1=0, \label{Integrability_conditions}
\end{eqnarray}
to make sure
that the associated $K$-matrices satisfy the RE (\ref{RE}) and its dual (\ref{DRE}).

Let us introduce the one-row monodromy matrices
\begin{eqnarray}
 &&T_0(u)=R_{0L}(u)R_{0L-1}(u)\cdots R_{01}(u), \\
 &&\hat{T}_0(u)=R_{1,0}(u)\cdots R_{L-1,0}(u)R_{L,0}(u),
\end{eqnarray}
and the double-row monodromy matrix
\begin{eqnarray}
 \mathcal{U}(u)=T(u)K^{-}(u)\hat{T}(u).
\end{eqnarray}
The transfer matrix  is given by
\begin{eqnarray}
 t(u)=\mathrm{str_0}\{ K_0^{+}(u)\mathcal{U}_0(u) \},
\end{eqnarray}
where $\mathrm{str_0}$ denotes the supertrace carried out in auxiliary space \cite{PhysRevB.46.9147, PhysRevB.61.3450}.

With the same procedure introduced in \cite{Skl88}, one can show that  $[t(u),t(v)]=0$, which ensures the
integrability of the model described by the Hamiltonian (\ref{Hamiltonian_basic}). The first order derivative of the logarithm of the transfer matrix
$t(u)$ yields the Hamiltonian (\ref{Hamiltonian_basic})
\begin{eqnarray}
H=\frac{\eta}{2} \frac{d\ln t(u)}{d u}\Big|_{u=0}-\frac{\eta}{2\xi}+\frac{\eta-2\xi'}{2(\eta-\xi')}+2\hat{N}-L+1, \label{transfer_H}
\end{eqnarray}
where the coupling constants in the Hamiltonian are expressed in terms of the parameters in the corresponding $K$-matrices
given in (\ref{Kmatrix}), (\ref{dualKmatrix}) and (\ref{Integrability_conditions}) as follows:
\begin{eqnarray}
&&\chi_{1}=1-\frac{\eta}{2\xi}, \qquad h_{1}^{x}=\frac{\eta}{2\xi}(c_2+c_1), \qquad h_{1}^{y}=\frac{\eta}{2\xi}(c_2-c_1)i,  \qquad h_{1}^{z}=-\frac{\eta}{2\xi}c, \nonumber \\
&& \chi_{L}=1-\frac{\eta}{2(\eta-\xi')}, \qquad h_{L}^{x}=\frac{\eta}{2(\eta-\xi')}(c_2'+c_1'), \nonumber \\
&& h_{L}^{y}=\frac{\eta}{2(\eta-\xi')}(c_2'-c_1')i, \qquad h_{L}^{z}=-\frac{\eta}{2(\eta-\xi')}c'. \label{parameter_correspondences}
\end{eqnarray}
It is remarked that the total number  operator $\hat{N}$ is still a conserved charge for the model described by the Hamiltonian (\ref{Hamiltonian_basic}), i.e., $[H,\hat{N}]=0$.

By combining the algebraic Bethe ansatz and the off-diagonal Bethe ansatz \cite{JHEP.07.051}, the eigenvalues $\Lambda(u)$ of the transfer matrix $t(u)$ is given by an inhomogeneous $T-Q$ relation
\begin{eqnarray}
\Lambda(u)
&=&w_3(u)(\xi+u)(u+\eta)^{2L}\frac{Q(u-\eta)}{Q(u)}
-u^{2L}\bar{a}(u)\frac{Q(u-\eta)Q^{(1)}(u+\eta)}{Q(u)Q^{(1)}(u)} \nonumber\\
&&-u^{2L}\bar{d}(u)\frac{Q^{(1)}(u-\eta)}{Q^{(1)}(u)}
+2hu^{2L+1}(u-\frac{\eta}{2})
\frac{Q(u-\eta)}{Q^{(1)}(u)},  \label{T_Q_relation}
\end{eqnarray}
where
\begin{eqnarray}
&& w_3(u)=\xi'-u-\frac{\eta}{2u+\eta}\left(2\xi'-\eta\right), \qquad h=1+(cc'+2c_1c_2'+2c_1'c_2),
\nonumber \\
&& \bar{a}(u)=\frac{u-\frac{\eta}{2}}{u+\frac{\eta}{2}}(u+\xi')(u+\xi),\qquad
\bar{d}(u)=(u-\xi')(u-\xi),\nonumber \\
&& Q^{(1)}(u)=\prod_{l=1}^{M}(u-\lambda_l)(u+\lambda_l),\qquad
Q(u)=\prod_{k=1}^{M}(u-v_k)(u+v_k+\eta). \label{non_diag_h}
\end{eqnarray}

For simplicity, we introduce the new parameters $\theta$ and $\varphi$ which satisfy
\begin{eqnarray}
c=\cos(\theta),~c_1=\frac{\sin(\theta)}{2}e^{i\varphi},~c_2=\frac{\sin(\theta)}{2}e^{-i\varphi}, \label{integrable_condition_01}
\end{eqnarray}
and other two new parameters $\theta'$ and $\varphi'$ which satisfy
\begin{eqnarray}
 c'=-\varepsilon\cos(\theta'),~c_1'=-\varepsilon\frac{\sin(\theta')}{2}e^{i\varphi'},~c_2'=-\varepsilon\frac{\sin(\theta')}{2}e^{-i\varphi'},  \label{integrable_condition_02}
\end{eqnarray}
where
\begin{eqnarray}
\varepsilon=\mathrm{sgn}(\mathbf{h}_{1}\cdot \mathbf{h}_{L})=\frac{\mathbf{h}_{1}\cdot \mathbf{h}_{L}}{|\mathbf{h}_{1}\cdot \mathbf{h}_{L}|}.
\end{eqnarray}
The above parameterizations make the constraints (\ref{Integrability_conditions}) fulfilled automatically. We further assume the parameters $\eta$, $\xi,\theta,\phi$, $\xi',\theta',\phi'$ being  real numbers to ensure the hermitian of the  Hamiltonian (\ref{Hamiltonian_basic}).
For $\varepsilon=1$ case, the possible taking values of the parameters $\xi$ and $\xi'$ are constrained in the region of $\xi<0$ and $\xi'<1$  or $\xi>0$ and $\xi'>1$ , respectively. While for $\varepsilon=-1$ case, the possible taking values of the parameters $\xi$ and $\xi'$ are constrained in the region of $\xi<0$ and $\xi'>1$  or $\xi>0$ and $\xi'<1$, respectively.
In this paper, we choose  the region of $\xi<0$ and $\xi'<1$ as an example.
It is straightforward to extend the analysis below to other ranges of the fields.
\footnote{We note that the conclusions may not change in the other ranges of the fields.}

Using the relations (\ref{non_diag_h}) - (\ref{integrable_condition_02}), we obtain
\begin{eqnarray}
h=1-\varepsilon\left[\cos(\theta)\cos(\theta')+\sin(\theta)\sin(\theta')\cos(\varphi-\varphi')\right]. \label{non_diag_h2}
\end{eqnarray}
It should be remarked that if the two boundary fields $\mathbf{h}_{1}$ and $\mathbf{h}_{L}$ are parallel (i.e., $\theta'=\theta$, $\varphi'=\varphi$) or anti-parallel (i.e., $\theta'+\theta=\pi$, $|\varphi'-\varphi|=\pi$), the associated $K^{\pm}$-matrices can be diagonalized simultaneously. In this case, the $U(1)$ symmetry in the spin sector is recovered and the constant $h$ given by  (\ref{non_diag_h2})  vanishes.

To ensure $\Lambda(u)$ to be a polynomial, the residues of $\Lambda(u)$ at the poles $v_j$ and $\lambda_j$ must  vanish, i.e., the $2M$ parameters $\{v_j|j=1,\cdots,M\}$ and $\{\lambda_j|j=1,\cdots,M\}$ must satisfy the nested Bethe ansatz equations (BAEs)
\begin{eqnarray}
\left(\xi'\hspace{-0.12truecm}-\hspace{-0.12truecm}v_j\hspace{-0.12truecm}-\hspace{-0.12truecm}\frac{\eta}{2v_j\hspace{-0.12truecm}+\hspace{-0.12truecm}\eta}
\left(2\xi'-\eta\right)\right)(\xi\hspace{-0.12truecm}+\hspace{-0.12truecm}v_j)(v_j\hspace{-0.12truecm}+\hspace{-0.12truecm}\eta)^{2L}=
v_j^{2L}\bar{a}(v_j)\frac{Q^{(1)}(v_j+\eta)}{Q^{(1)}(v_j)}, \,\, j=1,\cdots,M,\label{BAE1a}
\end{eqnarray}
and
\begin{eqnarray}
&&\hspace{-1.2truecm}\bar{a}(\lambda_j)Q(\lambda_j-\eta)Q^{(1)}(\lambda_j+\eta)
+\bar{d}(\lambda_j)Q(\lambda_j)Q^{(1)}(\lambda_j-\eta) \nonumber\\
&&\qquad\qquad\qquad\qquad=2h\lambda_j\left(\lambda_j-\frac{\eta}{2}\right)Q(\lambda_j)Q(\lambda_j-\eta),\quad j=1,\cdots,M. \label{BAE1b}
\end{eqnarray}
\begin{table}
\caption{Solutions of BAEs (\ref{BAE1a})-(\ref{BAE1b}) for the case of $L=2, \eta=1, \xi=0.6, \theta=\pi/5, \phi=\pi/3, \xi'=1.5, \theta'=2\pi/3, \phi'=\pi/4$. The symbol $n$ indicates the number of the eigenvalues, and $E_{n}$ is the corresponding eigenenergy. The energy $E_{n}$ calculated from (\ref{Energy1}) is the same as that from the exact diagonalization of the Hamiltonian (\ref{Hamiltonian_basic}).}
\centering 
\begin{small}
\begin{tabular}{|cccc|c|c|}
\hline $v_1$ & $v_2$ & $\lambda_1$ & $\lambda_2$ & $E_{n}$ & $n$  \\ \hline
$-0.5000-0.2786i$  &  $ --$  &  $0.4756+0.0000i$ &  $ --$  & $-1.052578$ & $1$  \\
$--$  &  $ --$  &  $--$ &  $ --$  & $0$ & $2$  \\
 $-0.5000-0.5801i$ &  $ --$ &  $1.5481-0.0000i$ &  $ --$ & $0.295101$ & $3$  \\
$-0.5000+0.2731i$  &  $-0.5000+1.6511i$  &  $-0.4393-0.0000i$  &  $0.0000+1.6139i$ & $0.583040$ & $4$  \\
 $-0.5000+1.9038i$  & $--$  &  $0.0000+2.1272i$ & $--$  &  $1.741892$ & $5$  \\
$-0.0015+0.3109i$  &  $-0.0015-0.3109i$  &  $-0.0000+0.3403i$  &  $2.1227-0.0000i$ & $2.141851$ & $6$  \\
$-0.5000-0.4606i$  &  $0.5419-0.0000i$  &  $0.9406+0.0000i$  &  $2.1803-0.0000i$ & $3.033107$ & $7$  \\
 $0.4957-0.0000i$  & $--$  &  $1.6356+0.0000i$ & $--$  &  $3.348918$ & $8$  \\
$0.4798-0.0000i$  &  $-0.5000-1.3235i$  &  $-0.0000+1.5362i$  &  $2.0858+0.0000i$ & $4.908668$ & $9$  \\
 \hline\end{tabular}  \label{tab:BAEs}
\end{small}
\end{table}
From the relation (\ref{transfer_H}), we have the eigenvalue of the Hamiltonian (\ref{Hamiltonian_basic}) in terms of the
Bethe roots, which is given by
\begin{eqnarray}
E&=&\frac{\eta}{2} \frac{d\ln \Lambda(u)}{d u}\Big|_{u=0}-\frac{\eta}{2\xi}+\frac{\eta-2\xi'}{2(\eta-\xi')}+2M-L+1 \nonumber\\
&=&\sum_{k=1}^{M}\frac{\eta^2}{v_k(v_k+\eta)}+2M. \label{Energy1}
\end{eqnarray}
\begin{figure}[ht]
\centering
\includegraphics[width=4.0in]{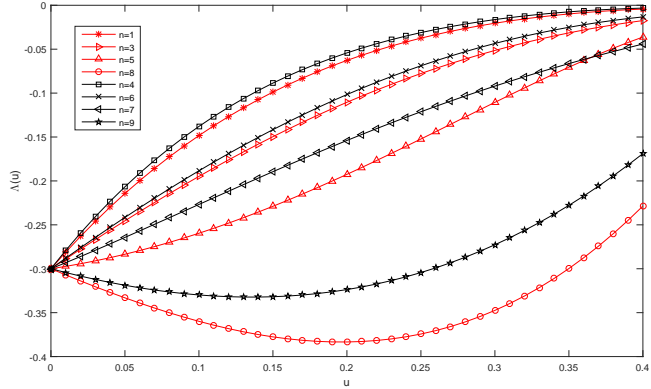}
\caption{(color online) $\Lambda(u)$ vs. $u$ for the case of  $L=2$. The curves calculated from $T-Q$ relation (\ref{T_Q_relation}) and the nested BAEs (\ref{BAE1a})-(\ref{BAE1b}) are exactly the same as those obtained from the exact diagonalization of the transfer matrix $t(u)$.}\label{Lambda_u}
\end{figure}

The numerical solutions of BAEs (\ref{BAE1a})-(\ref{BAE1b}) and the corresponding eigenvalues of the Hamiltonian (\ref{Hamiltonian_basic}) for $L=2$ is shown in Table \ref{tab:BAEs}, while the calculated $\Lambda(u)$
curves for $L=2$ are shown in Figure \ref{Lambda_u}. Those numerical simulations imply that the inhomogeneous $T-Q$
relation (\ref{T_Q_relation}) and the BAEs (\ref{BAE1a})-(\ref{BAE1b}) indeed give the correct and complete spectrum of
the one-dimensional supersymmetric $t-J$ model with unparallel boundary fields \cite{wang2015off, J.Phys.A.46.442002, J.Stat.Mech.1502.P02001}.


\section{Finite-size effects}
\label{inhomogeneous term} \setcounter{equation}{0}
In order to study the contribution of the inhomogeneous term (the last term in  (\ref{T_Q_relation})) to the ground state energy,
we first consider the $T-Q$
relation without the inhomogeneous term\footnote{It should be emphasized that, for a finite $L$, $\Lambda_{hom}(u)$
is different from the exact eigenvalue $\Lambda(u)$ given by (\ref{T_Q_relation}).}, i.e.,
\begin{eqnarray}
\Lambda_{hom}(u)
&=&w_3(u)(\xi+u)(u+\eta)^{2L}\frac{Q(u-\eta)}{Q(u)}
-u^{2L}\bar{a}(u)\frac{Q(u-\eta)Q^{(1)}(u+\eta)}{Q(u)Q^{(1)}(u)} \nonumber\\
&&-u^{2L}\bar{d}(u)\frac{Q^{(1)}(u-\eta)}{Q^{(1)}(u)}.  \label{T_Q_relation_homo}
\end{eqnarray}
The singular property of the $T-Q$ relation (\ref{T_Q_relation_homo}) gives rise to  the associated BAEs
\begin{eqnarray}
\frac{\mu_j+(\frac{\eta}{2}-\xi')}{\mu_j-(\frac{\eta}{2}-\xi')}\left(\frac{\mu_j+\frac{\eta}{2}}{\mu_j-\frac{\eta}{2}}\right)^{2L}
=-\prod_{l=1}^{\bar{M}}\frac{\mu_j-\lambda_l+\frac{\eta}{2}}{\mu_j-\lambda_l-\frac{\eta}{2}}
\frac{\mu_j+\lambda_l+\frac{\eta}{2}}{\mu_j+\lambda_l-\frac{\eta}{2}},
\end{eqnarray}
and
\begin{eqnarray}
&&\frac{\lambda_j-\frac{\eta}{2}}{\lambda_j+\frac{\eta}{2}}\frac{(\lambda_j+\xi')(\lambda_j+\xi)}{(\lambda_j-\xi')(\lambda_j-\xi)}
\nonumber\\
&&=-\prod_{k=1}^{M}\frac{(\lambda_j-\mu_k+\frac{\eta}{2})(\lambda_j+\mu_k+\frac{\eta}{2})}
{(\lambda_j-\mu_k-\frac{\eta}{2})(\lambda_j+\mu_k-\frac{\eta}{2})}
\prod_{l=1}^{\bar{M}}\frac{(\lambda_j-\lambda_l-\eta)(\lambda_j+\lambda_l-\eta)}{(\lambda_j-\lambda_l+\eta)(\lambda_j+\lambda_l+\eta)},
\end{eqnarray}
where we have put $v_j=\mu_j-\frac{\eta}{2}$.

Assume that $\mu_j\rightarrow \mu_j i$, $\lambda_j\rightarrow \lambda_j i$ and $\eta=1$, we obtain
\begin{eqnarray}
\frac{\mu_j-(\frac{1}{2}-\xi') i}{\mu_j+(\frac{1}{2}-\xi' )i}\left(\frac{\mu_j-\frac{i}{2}}{\mu_j+\frac{i}{2}}\right)^{2L}
=-\prod_{l=1}^{\bar{M}}\frac{\mu_j-\lambda_l-\frac{i}{2}}{\mu_j-\lambda_l+\frac{i}{2}}
\frac{\mu_j+\lambda_l-\frac{i}{2}}{\mu_j+\lambda_l+\frac{i}{2}}, \label{BAEs2a}
\end{eqnarray}
and
\begin{eqnarray}
&&\frac{\lambda_j+\frac{i}{2}}{\lambda_j-\frac{i}{2}}\frac{(\lambda_j-\xi' i)(\lambda_j-\xi i)}{(\lambda_j+\xi' i)(\lambda_j+\xi i)} \nonumber\\
&&=-\prod_{k=1}^{M}\frac{(\lambda_j-\mu_k-\frac{i}{2})(\lambda_j+\mu_k-\frac{i}{2})}
{(\lambda_j-\mu_k+\frac{i}{2})(\lambda_j+\mu_k+\frac{i}{2})}
\prod_{l=1}^{\bar{M}}\frac{(\lambda_j-\lambda_l+i)(\lambda_j+\lambda_l+i)}{(\lambda_j-\lambda_l-i)(\lambda_j+\lambda_l-i)}. \label{BAEs2b}
\end{eqnarray}
The corresponding eigenvalue reads
\begin{eqnarray}
E_{hom}&=&\frac{\eta}{2} \frac{d\ln \Lambda_{hom}(u)}{d u}\Big|_{u=0}-\frac{\eta}{2\xi}+\frac{\eta-2\xi'}{2(\eta-\xi')}+2M-L+1 \nonumber\\
&=&-\sum_{k=1}^{M}\frac{1}{\mu_k^2+\frac{1}{4}}+2M. \label{Energy2}
\end{eqnarray}

Now, we consider the contribution of the inhomogeneous term in Eq. (\ref{T_Q_relation}) to the ground state energy of the system.
In order to this, we should analyze the distribution of Bethe roots in the BAEs (\ref{BAEs2a}) and (\ref{BAEs2b}).
For $\xi<0$ and $\xi'<1$ (equivalent to $\chi_{1}>1$ and $\chi_{L}<1$), by using the standard thermodynamic Bethe ansatz method and taking the limit of temperature tending to zero, we find that all the Bethe roots are real at the ground state in the region of $\xi<0$ and $\xi'\leq1/2$ (equivalent to $\chi_{1}>1$ and $0\leq\chi_{L}<1$).
While in region of $\xi<0$ and $1/2<\xi'<1$ (equivalent to $\chi_{1}>1$ and $\chi_{L}<0$), besides the real Bethe roots, there exists an imaginary Bethe root which corresponds to a boundary bound state.
Let us discuss them separately.

\subsection{Region of $\xi<0$ and $\xi'\leq1/2$}

Firstly, we consider the case of
$\xi<0$ and $\xi'\leq1/2$ \cite{J.Phys.A:Math.Gen.29.6183, wang2015off}, in which all the Bethe roots are real at the ground state.
Taking the logarithm of BAEs (\ref{BAEs2a})-(\ref{BAEs2b}), we obtain
\begin{eqnarray}
2\pi I_{j}&=&2\arctan\left(\frac{2\mu_j}{1-2\xi'}\right)+4L\arctan(2\mu_j)  \nonumber\\
&&-\sum_{l=1}^{\bar{M}}2\arctan(2(\mu_j-\lambda_l))+2\arctan(2(\mu_j+\lambda_l)),
\label{BAEs2alog}\\
2\pi J_{j}&=&2\arctan(2\lambda_j)-2\arctan\left(\frac{\lambda_j}{\xi'}\right)-2\arctan\left(\frac{\lambda_j}{\xi}\right) \nonumber\\
&&+\sum_{k=1}^{M}2\arctan(2(\lambda_j-\mu_k))+2\arctan(2(\lambda_j+\mu_k))  \nonumber\\
&&-\sum_{l=1}^{\bar{M}}2\arctan(\lambda_j-\lambda_l)+2\arctan(\lambda_j+\lambda_l), \label{BAEs2blog}
\end{eqnarray}
where $I_{j}$ and $J_{j}$ are both quantum numbers which determine
the eigenenergy and the corresponding eigenstates.
It is well-known
that the size of the system $L$, with either even or odd value, gives the same
physics properties in the thermodynamic limit. Therefore, for simplicity, we set $L$ as an even
number.

We define the contribution of the inhomogeneous term to the ground state energy as
\begin{eqnarray}\label{DeltaE}
  E_{inh}=E_{hom}-E_{true}.
\end{eqnarray}
Here $E_{hom}$ is the energy of the supersymmetric $t-J$ model calculated by the eigenvalue (\ref{Energy2}) and the BAEs (\ref{BAEs2alog})-(\ref{BAEs2blog}).
$E_{true}$ is the energy of the
Hamiltonian (\ref{Hamiltonian_basic}), which can be  obtained  by   using the density matrix renormalization group (DMRG)
\cite{JSMTE.05.05001}.
For the ground state, the number of Bethe roots reduces to $M=L/2$ and $\bar{M}=0$,
\begin{eqnarray}
2\pi I_{j}=2\arctan\left(\frac{\mu_j}{\zeta}\right)+4L\arctan(2\mu_j),~j=1,\cdots,M, \label{BAEs3}
\end{eqnarray}
where $I_j \in \{1,2,\cdots,L/2\}$, $\zeta=1/2-\xi'$ and $\zeta\geq0$.
Then ``ground state energy"
$E_{hom}$ is given by equation (\ref{Energy2}) with the constraint
(\ref{BAEs3}).

\begin{figure}[ht]
\centering
\subfigure[$\theta=\pi/5$]{
\begin{minipage}{5cm}
\centering
\includegraphics[scale=0.448]{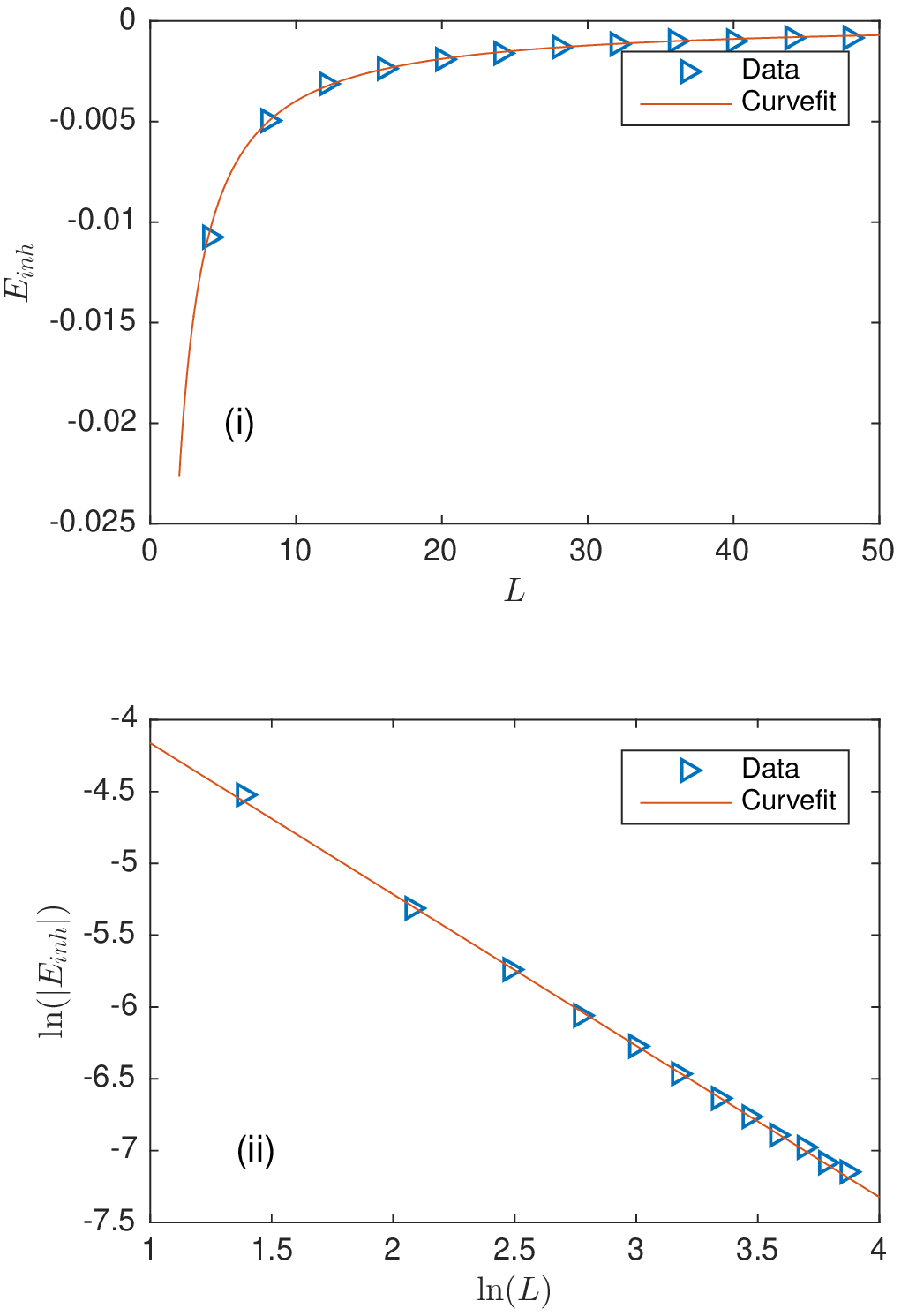}
\end{minipage}}
\subfigure[$\theta=\pi/4$]{
\begin{minipage}{5cm}
\centering
\includegraphics[scale=0.448]{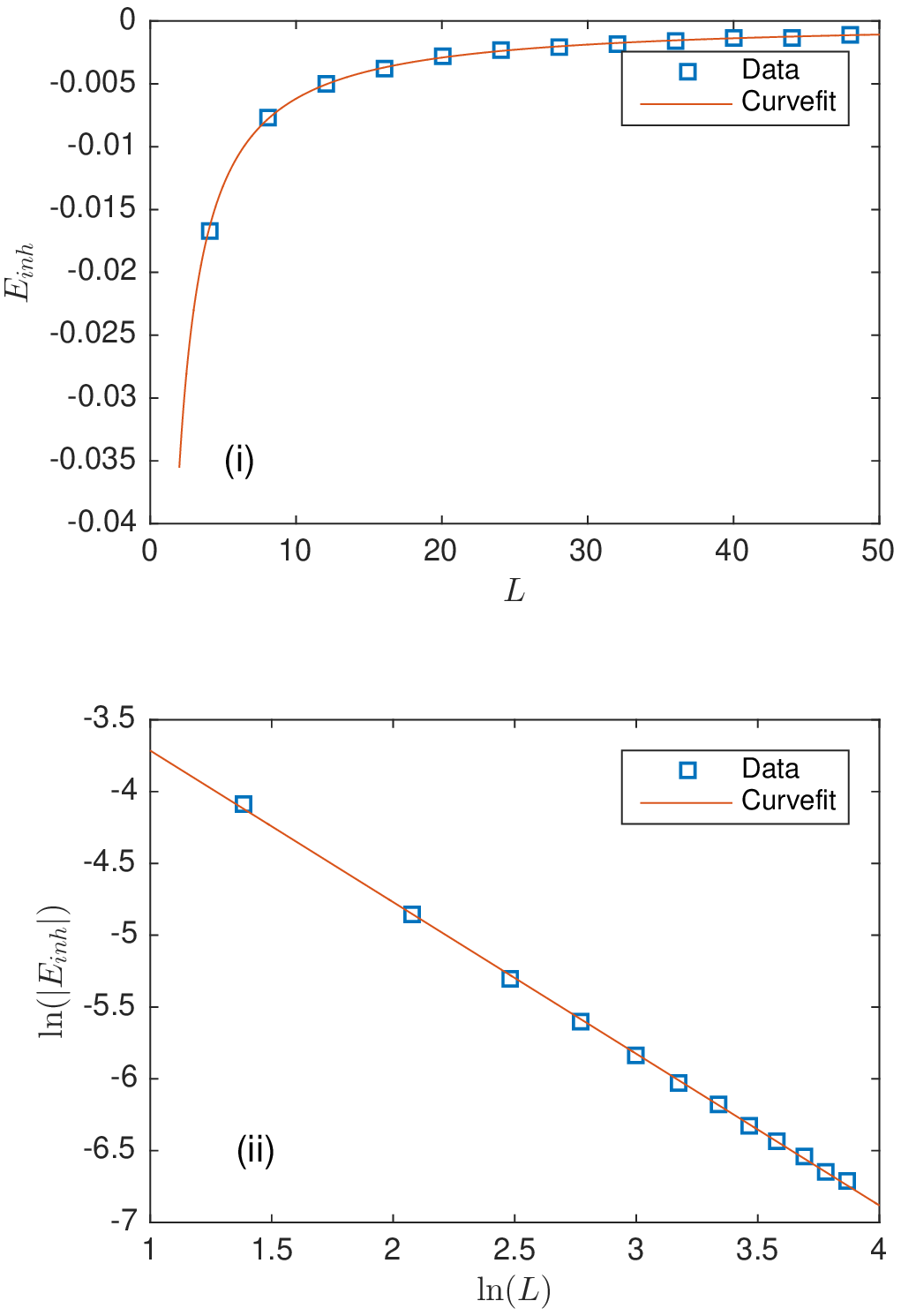}
\end{minipage}}
\subfigure[$\theta=\pi/2$]{
\begin{minipage}{5cm}
\centering
\includegraphics[scale=0.448]{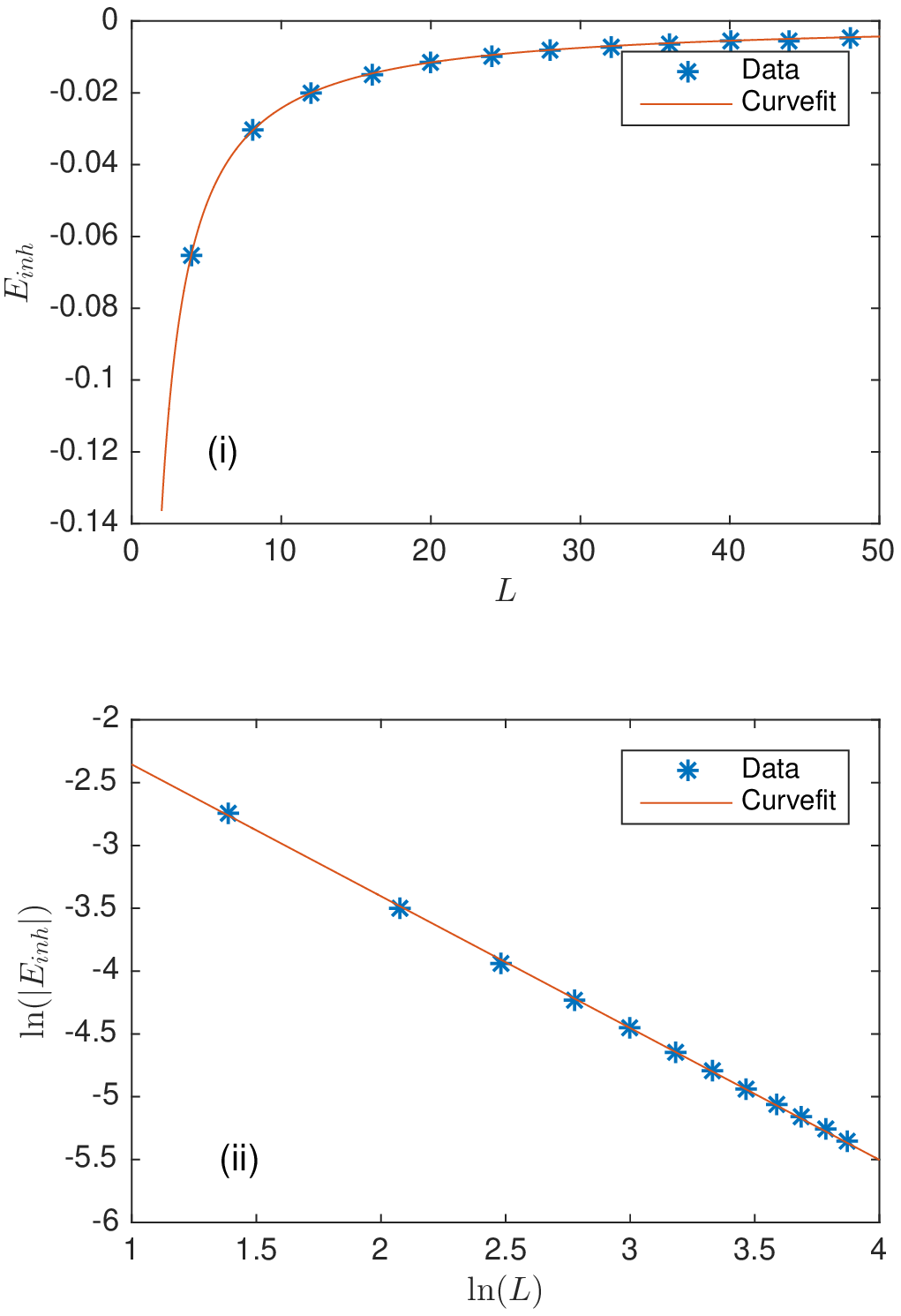}
\end{minipage}}
\caption{The contribution of the inhomogeneous term to the ground state energy
$E_{inh}$ versus the system size $L$. The data can be fitted as
$E_{inh}=\gamma L^{\beta}$ [or $\ln(|E_{inh}|)=p \ln(L)+q$]. Here $\xi=-0.1$, $\zeta=0.05$, $\theta'=0$, $\phi=\phi'=0$,
(a) $\theta=\pi/5$, $\gamma=-0.0478$ and $\beta=-1.080$;
(b) $\theta=\pi/4$, $\gamma=-0.0755$ and $\beta=-1.087$;
(c) $\theta=\pi/2$, $\gamma=-0.2871$ and $\beta=-1.073$. Due to the fact $\beta<0$ [or $p<0$],
when the $L$ tends to infinity, the contribution of the
inhomogeneous term tends to zero. }
\label{fig:2}
\end{figure}

The values of $E_{inh}$, the contribution of the inhomogeneous term to the ground state energy, versus the system size $L$ are shown in Figure \ref{fig:2}.
From the fitting, we find the power law relation
between $E_{inh}$ and $L$, i.e., $E_{inh}=\gamma L^{\beta}$.
Due to the fact that $\beta\approx-1$, the value of $E_{inh}$ tends to zero when
the size of the system tends to infinity, which means that the
inhomogeneous term in the $T-Q$ relation (\ref{T_Q_relation}) can be neglected in the
thermodynamic limit $L, N, M \rightarrow \infty$ with $N/L$ and $M/L$ kept fixed.
Therefore, the two boundaries are decoupled from each other completely in the
thermodynamic limit.
When $h=0$, the unparallel boundary fields degenerates into the parallel one.
At this point,  the contribution of the inhomogeneous term to the ground state energy $E_{inh}$ is equal to 0.

\subsection{Region of $\xi<0$ and $1/2<\xi'<1$}

\begin{figure}[ht]
\centering
\subfigure[$\theta'=0.80$]{
\begin{minipage}{5cm}
\centering
\includegraphics[scale=0.448]{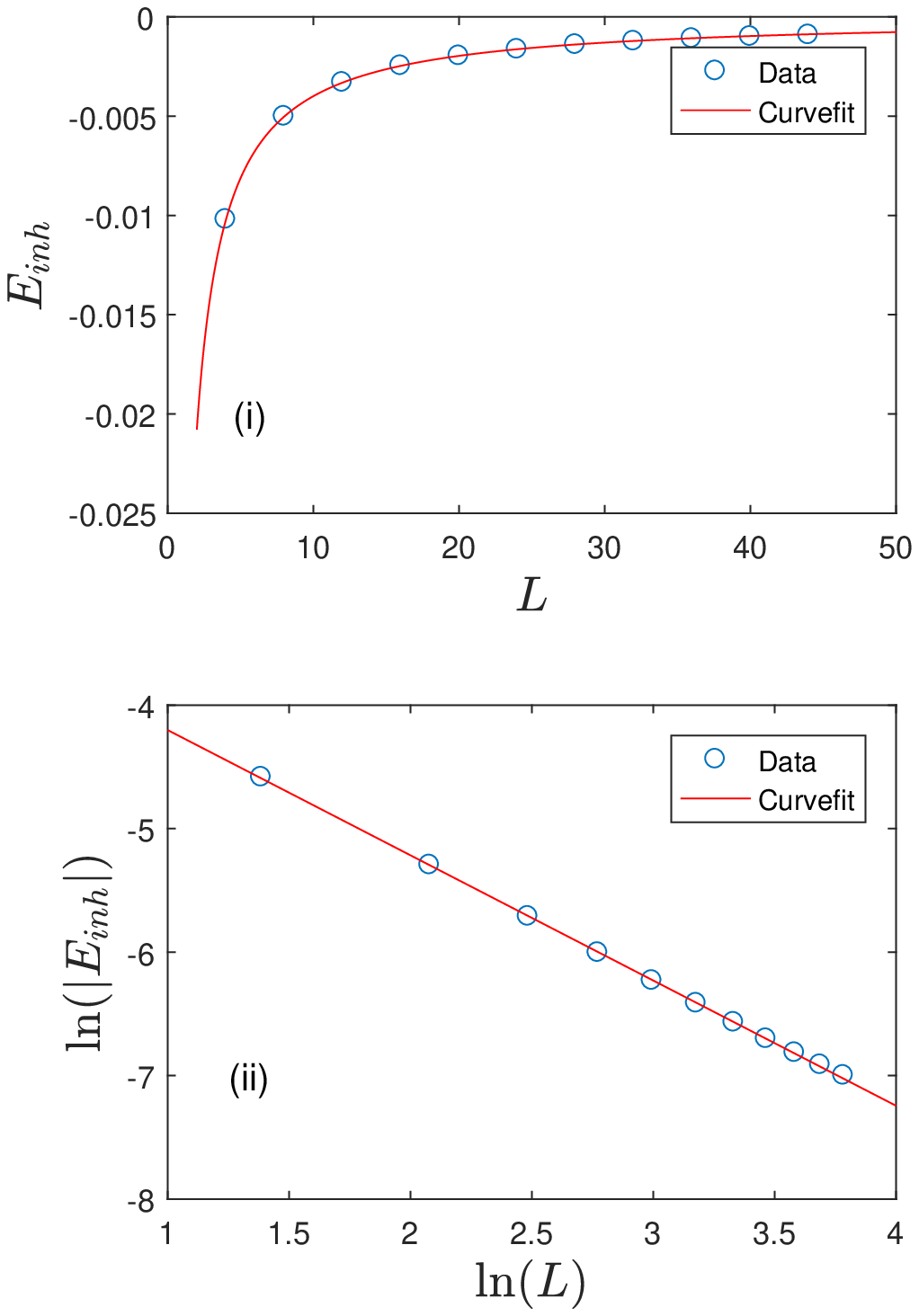}
\end{minipage}}
\subfigure[$\theta'=1.20$]{
\begin{minipage}{5cm}
\centering
\includegraphics[scale=0.448]{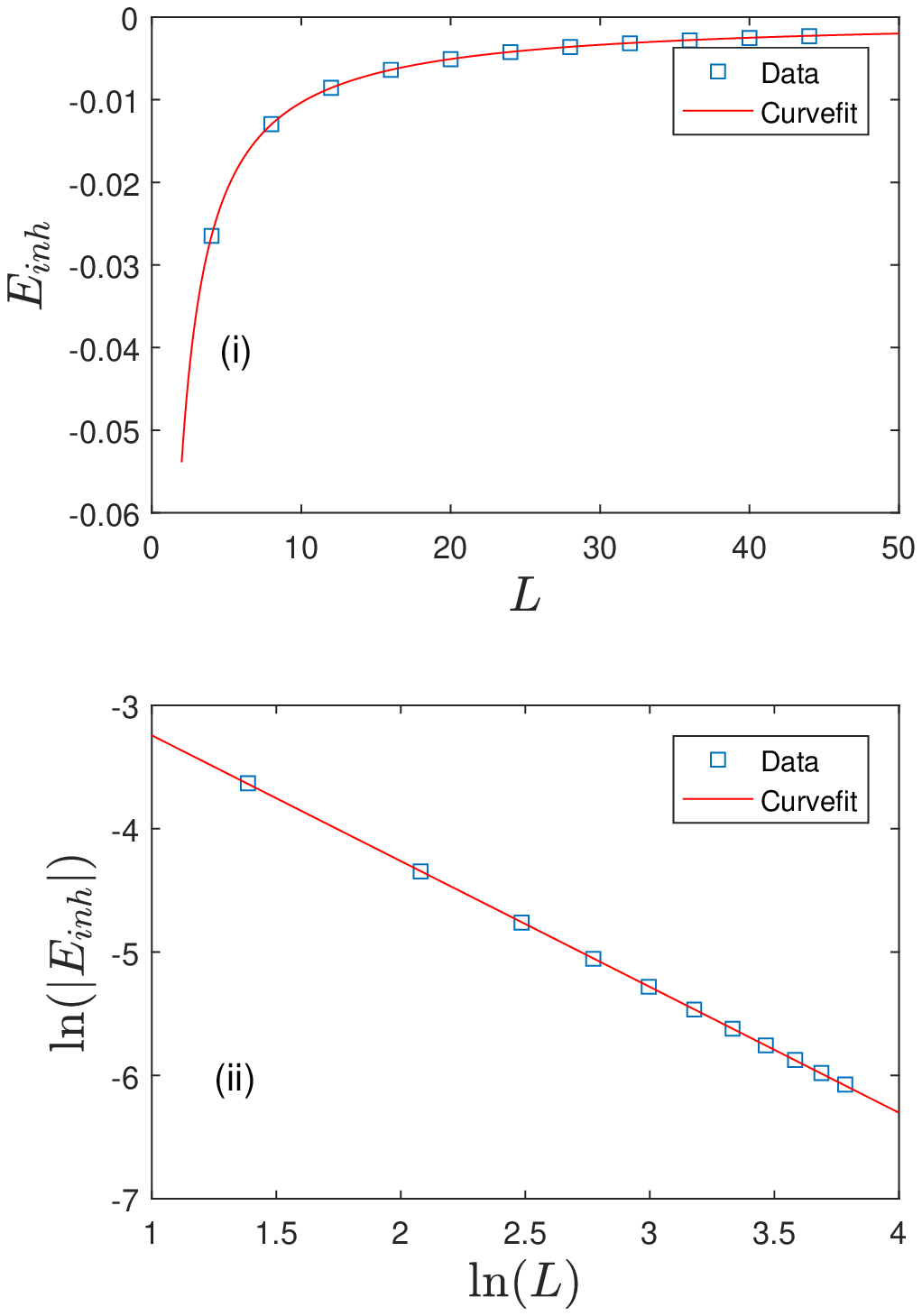}
\end{minipage}}
\subfigure[$\theta'=1.57$]{
\begin{minipage}{5cm}
\centering
\includegraphics[scale=0.448]{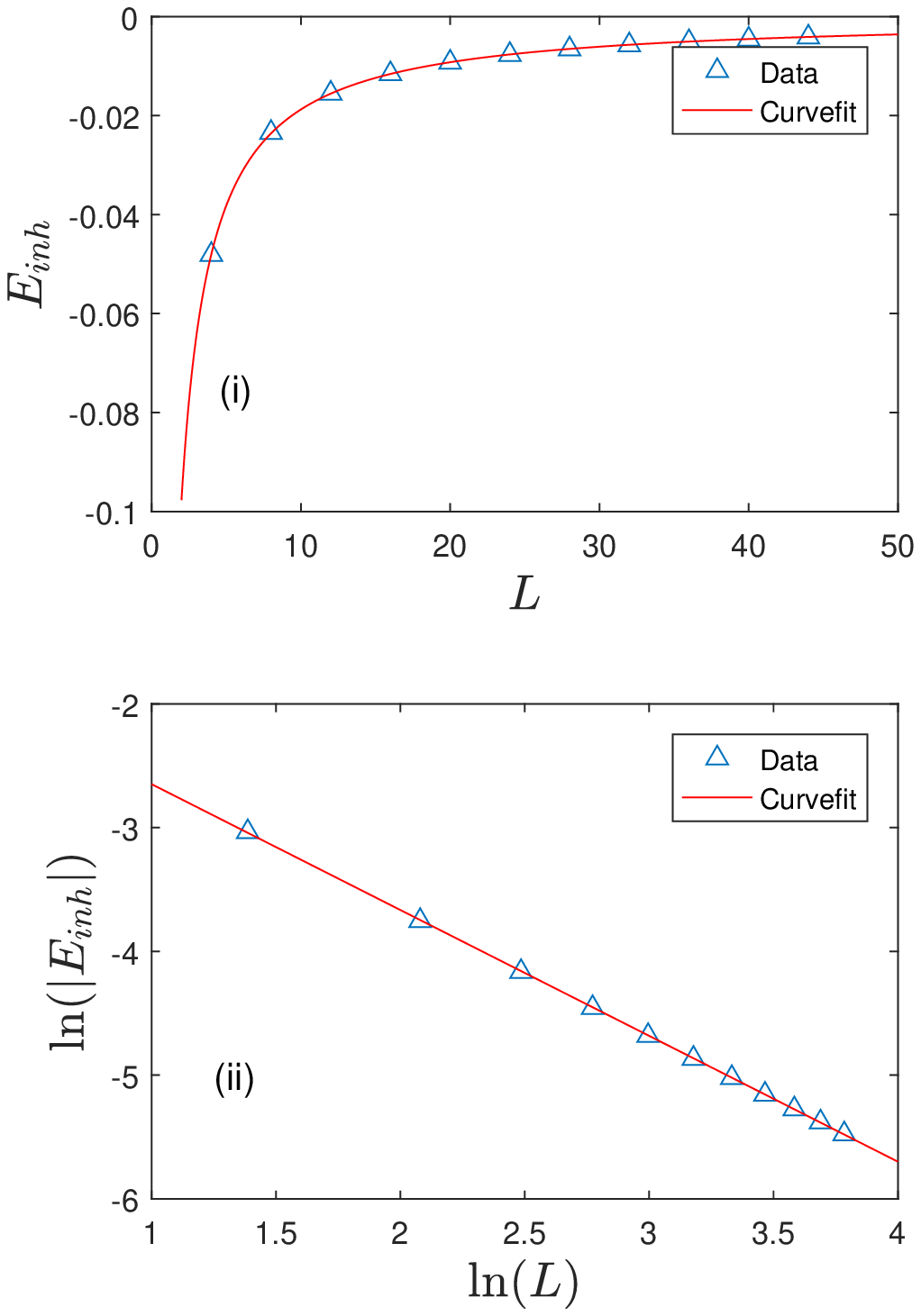}
\end{minipage}}
\caption{The values of $E_{inh}$ versus the system size $L$. The data can be fitted as
$E_{inh}=\tilde{\gamma} L^{\tilde{\beta}}$ [or $\ln(|E_{inh}|)=\tilde{p} \ln(L)+\tilde{q}$]. Here $\xi=-0.1$, $\xi'=0.95$, $\theta=0.15$, $\phi=\phi'=0$,
(a) $\theta'=0.80$, $\tilde{\gamma}=0.0423$ and $\tilde{\beta}=-1.0250$;
(b) $\theta'=1.20$, $\tilde{\gamma}=0.1097$ and $\tilde{\beta}=-1.0260$;
(c) $\theta'=1.57$, $\tilde{\gamma}=0.1988$ and $\tilde{\beta}=-1.0250$. Due to the fact $\tilde{\beta}<0$ [or $\tilde{p}<0$],
when the $L$ tends to infinity, the $E_{inh}$ tends to zero. }
\label{fig:3}
\end{figure}

In the region of $\xi<0$ and $1/2<\xi'<1$,
one of the Bethe roots at the ground state goes to $(\frac{1}{2}-\xi')i$ when the system-size $L$ tends to infinity \cite{wang2015off, J.Phys.A.28.6605, Phys.Rev.B54.8491, Phys.Rev.B56.14045}. We note the value of this Bethe root is related with the boundary parameter $\xi'$. Without losing generality, we assume that $\mu_{M}=(\frac{1}{2}-\xi')i+\mathcal{O}(e^{-\delta L})=\zeta i+\mathcal{O}(e^{-\delta L})$
where $M=L/2$ and $\delta$ is a small positive number to account for the finite size
deviations. This Bethe root contributes a negative bare energy if $-1/2<\zeta<0$.
The remaining Bethe roots should take real values and satisfy the following BAEs
\begin{eqnarray}
\frac{\mu_j-\zeta i}{\mu_j+\zeta i}\left(\frac{\mu_j-\frac{i}{2}}{\mu_j+\frac{i}{2}}\right)^{2L}
=-1,~~~j=1,2,\cdots,M-1. \label{BS_BAEs}
\end{eqnarray}
Taking the logarithm of Eq.(\ref{BS_BAEs}), we have
\begin{eqnarray}
2\pi I_{j}=2\arctan\left(\frac{\mu_j}{\zeta}\right)+4L\arctan(2\mu_j),~j=1,\cdots,M-1, \label{BS_BAEs_log}
\end{eqnarray}
where the quantum numbers $\{I_j\}$ are chosen as $\{1,2,\cdots,L/2-1\}$.
The corresponding energy reads
\begin{eqnarray}
E_{hom}=-\sum_{k=1}^{M-1}\frac{1}{\mu_k^2+\frac{1}{4}}-\frac{1}{\frac{1}{4}-\zeta^2}+2M. \label{BS_E}
\end{eqnarray}

The values of $E_{inh}$ versus the system size $L$ are shown in Figure \ref{fig:3}.
From the fitted curves in Figure \ref{fig:3}, we see that the $E_{inh}$ also satisfies the scaling law $L^{-1}$.
\footnote{It should be emphasized that, for a finite $L$, the causes of the difference $E_{inh}$ included two aspects, omitting the exponentially small corrections and ignoring the contribution of the inhomogeneous term.}
Therefore,
the contribution of the inhomogeneous term to the ground state energy in the thermodynamic limit is zero and we have $E_{hom}=E_{ture}\equiv E$.
In addition, the results indicating that the boundary bound state should be stable. The surface energy will compute in the next section.

\section{Surface energy}
\label{Surface energy} \setcounter{equation}{0}

In order to analyze the influence of the boundary fields, now we calculate the surface energy \cite{PhysRevA.4.386,
JournalofPhysicsA.20.5677, JournalofPhysicsA.23.761} of the system.

\begin{figure}[ht]
  \centering
  \includegraphics[width=3.0in]{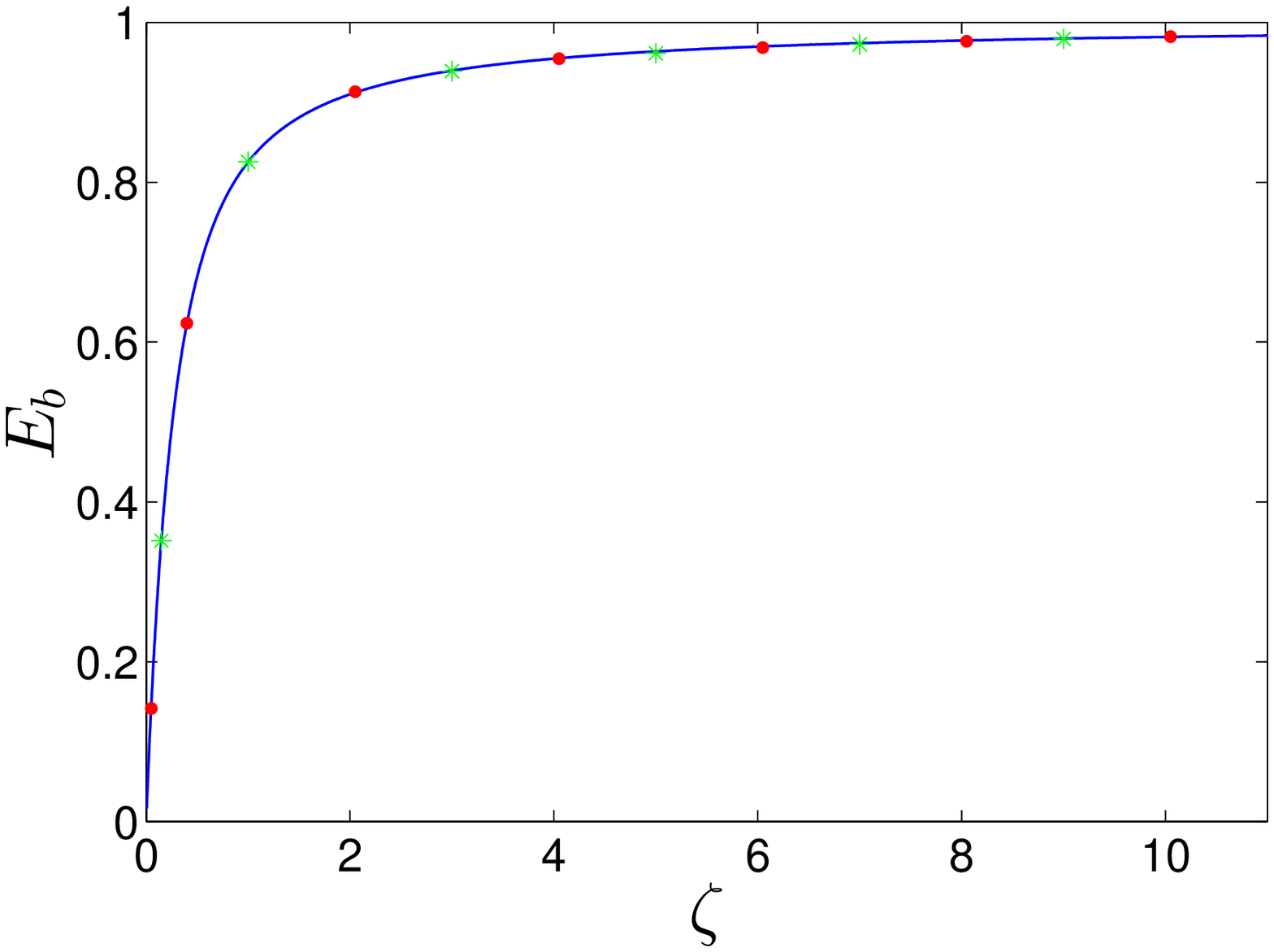}\\
  \caption{The surface energies versus the boundary parameters. The blue curves are the ones calculated from equation (\ref{TL_Eb}),
  while the red points ($\xi=-0.1$, $\theta=\pi/5$, $\phi=0$, $\theta'=0$ and $\phi'=0$) and green stars ($\xi=-5$, $\theta=\pi/4$, $\phi=0$, $\theta'=2\pi/3$ and $\phi'=0$) are the ones obtained from the Hamiltonian (\ref{Hamiltonian_basic}) with the BST algorithms.}\label{tJ-Eb_xipb}
\end{figure}

\subsection{Region of $\xi<0$ and $\xi'\leq1/2$}

Define $Z(\mu_{j})=\frac{I_{j}}{2L}$, then the BAEs (\ref{BAEs3}) can be rewritten as
\begin{eqnarray}
 Z(\mu_j)=\frac{1}{2\pi}\left[\frac{1}{2L}\Xi_{2\zeta}(\mu_j)+\Xi_{1}(\mu_j)\right],
\end{eqnarray}
where $\Xi_{n}(x)=2\arctan(2x/n)$.
It turns to be a continuous function in the thermodynamic limit as the distribution of Bethe roots is
continuous, i.e., $Z(\mu_{j}) \to Z(u)$.
In the thermodynamic limit, the density distributions are determined by
\begin{eqnarray}
\rho(u)+\rho^{h}(u)=\frac{d Z(u)}{d u}.
\end{eqnarray}
Taking the derivative of
$Z(u)$ with respect to $u$, we obtain the density of states as
\begin{eqnarray}
\rho(u)=a_{1}(u)+\frac{1}{2L}\left[a_{2\zeta}(u)-\delta(u)\right],
\end{eqnarray}
where
\begin{eqnarray}
  a_{n}(u)=\frac{1}{2\pi}\frac{n}{u^2+\frac{n^2}{4}}.
\end{eqnarray}
The ground state energy is equal to
\begin{eqnarray}
E=-2\pi L\int_{-B}^{B}\left(a_{1}(\mu)\right)^2d\mu+2M
-\pi\int_{-B}^{B}a_{1}(\mu)\left[a_{2\zeta}(\mu)-\delta(\mu)\right]d\mu,
\end{eqnarray}
and the energy density of the ground state is
\begin{eqnarray}
    e_{g}=-2\pi \int_{-B}^{B}\left(a_{1}(\mu)\right)^2d\mu+1+{O}(L^{-1}),
\end{eqnarray}
\begin{figure}[ht]
  \centering
  \includegraphics[width=3.0in]{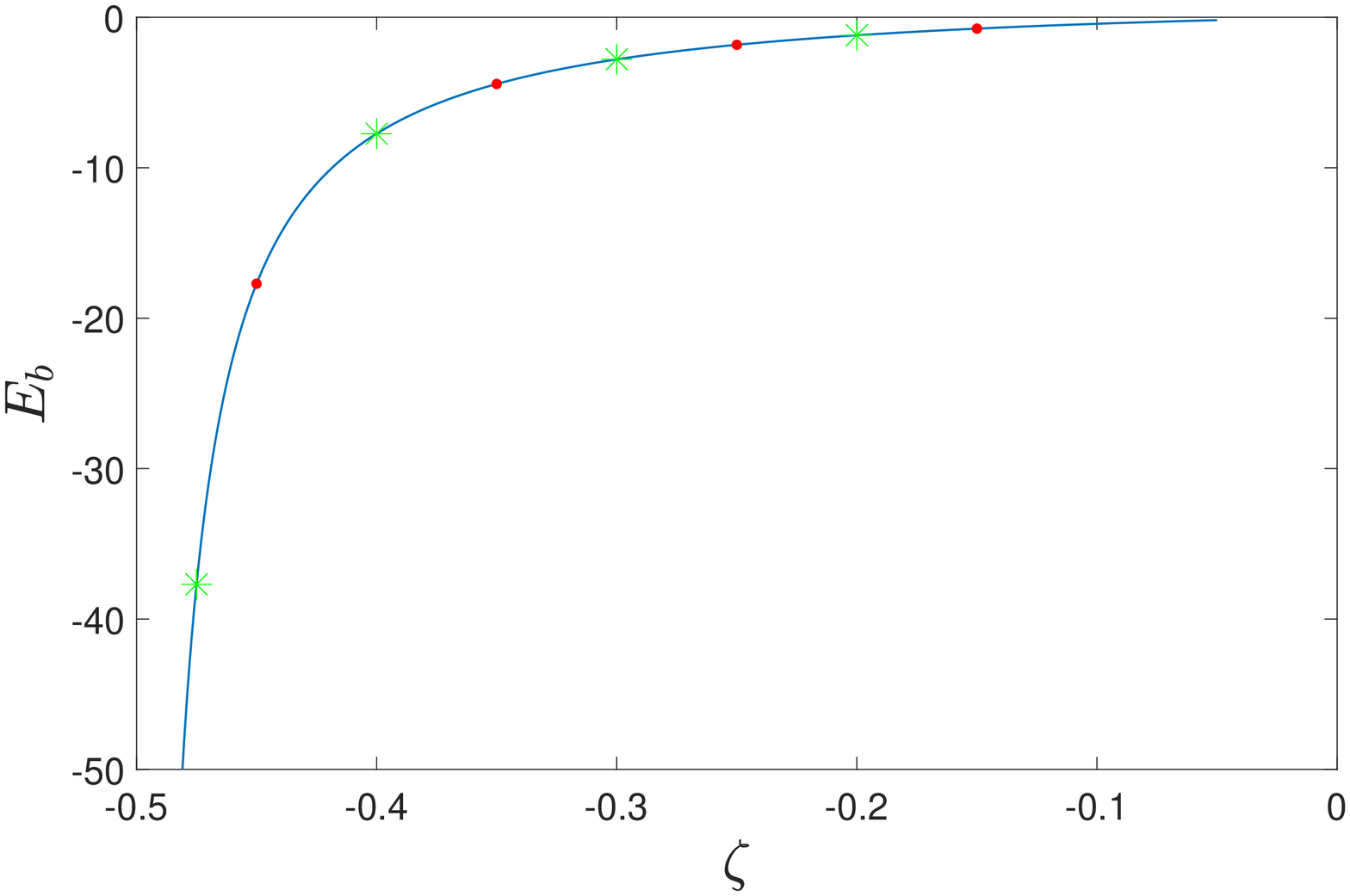}\\
  \caption{The surface energies versus the boundary parameters. The blue curves are the ones calculated from equation (\ref{tJ-Eb_zeta_BDS}),
  while the red points ($\xi=-0.1$, $\theta=0.3\pi$, $\phi=0$, $\theta'=0.7\pi$ and $\phi'=0$) and green stars ($\xi=-0.1$, $\theta=0.17\pi$, $\phi=0$, $\theta'=0.62\pi$ and $\phi'=0$) are the ones obtained from the Hamiltonian (\ref{Hamiltonian_basic}) with the BST algorithms.}\label{tJ-Eb_zeta}
\end{figure}
where $B=\frac{1}{2}+\frac{\pi-2\arctan\left(\frac{1}{2\zeta}\right)}{4L+\frac{4\zeta}{1+4\zeta^2}}$.
The energy density $e_{g}$ is equal to $-2/\pi$ in the thermodynamic limit, which is the same with that of the periodic case \cite{JMathPhys.15.1675}.
The surface energy then can be given by
\begin{eqnarray}
    &&E_{b}(\zeta)=\lim_{L\rightarrow\infty}\left[E_{0}(L;\zeta)-E_{0}^{periodic}(L)\right]\nonumber \\
    &&\quad\quad\quad=-\pi\int_{-\frac{1}{2}}^{\frac{1}{2}}a_{1}(\mu)a_{2\zeta}(\mu)d\mu+\frac{2}{\pi}\arctan\left(\frac{1}{2\zeta}\right)+1. \label{TL_Eb}
\end{eqnarray}
By using the relation (\ref{TL_Eb}), one can calculate the surface energy of the one-dimensional supersymmetric $t-J$ model with unparallel boundary fields. The results are shown
in Figure \ref{tJ-Eb_xipb}, where the blue solid lines are the surface energy calculated by using the relation (\ref{TL_Eb}) and
the red points and green stars are data obtained by employing the BST algorithms \cite{Journal.of.PhysicsA.21.11} to solve the surface energy of the Hamiltonian (\ref{Hamiltonian_basic})
in the thermodynamic limit. Specifically, for one of the red points or green stars,  we first calculate the ground state energy $E_{0}(L)$ with $L=4,8,\cdots,48$ by the DMRG.
Then, the large-$L$ extrapolation of the surface energy was performed using BST algorithms from the sequence $E_{0}(4)-4e_{g}^{\infty}$, $E_{0}(8)-8e_{g}^{\infty}$, $\cdots$, $E_{0}(48)-48e_{g}^{\infty}$. Note that $e_{g}^{\infty}=-2/\pi$.
From the Figure \ref{tJ-Eb_xipb}, we can see that the analytical and numerical results agree with each other very well for all tunable parameters. The surface energy increased with the increase of $\zeta$.
Taking the $\zeta\rightarrow0$ limit of Eq.(\ref{TL_Eb}), we have $E_{b}(\zeta\rightarrow0)=0$.
Taking the $\zeta\rightarrow\infty$ limit of Eq.(\ref{TL_Eb}), we have $E_{b}(\zeta\rightarrow\infty)=1$.

\subsection{Region of $\xi<0$ and $1/2<\xi'<1$}

In this region, the ground state energy of the system in the thermodynamic limit reads
\begin{eqnarray}
E=-2\pi L\int_{-\tilde{B}}^{\tilde{B}}\left(a_{1}(\mu)\right)^2d\mu+2M
-\pi\int_{-\tilde{B}}^{\tilde{B}}a_{1}(\mu)[a_{2\zeta}(\mu)-\delta(\mu)]d\mu-\frac{1}{\frac{1}{4}-\zeta^2},
\end{eqnarray}
and the energy density of the ground state is
\begin{eqnarray}
    e_{g}=-2\pi \int_{-\tilde{B}}^{\tilde{B}}\left(a_{1}(\mu)\right)^2d\mu+1+{O}(L^{-1}),
\end{eqnarray}
where $\tilde{B}=\frac{1}{2}-\frac{\pi+2\arctan\left(\frac{1}{2\zeta}\right)}{4L+\frac{4\zeta}{1+4\zeta^2}}$.
The surface energy is given by
\begin{eqnarray}
E_{b}(\zeta)
=-\pi\int_{-\frac{1}{2}}^{\frac{1}{2}}a_{1}(\mu)a_{2\zeta}(\mu)d\mu+\frac{2}{\pi}\arctan\left(\frac{1}{2\zeta}\right)+3-\frac{1}{\frac{1}{4}-\zeta^2}. \label{tJ-Eb_zeta_BDS}
\end{eqnarray}
The results are shown in Figure \ref{tJ-Eb_zeta}. Again, we see that the analytical results and the numerical ones agree with each other very well.

The surface energy can be written in unified forms as
\begin{eqnarray}
E_{b}(\zeta)
=-\pi\int_{-\frac{1}{2}}^{\frac{1}{2}}a_{1}(\mu)a_{2\zeta}(\mu)d\mu+\frac{2}{\pi}\arctan\left(\frac{1}{2\zeta}\right)+\Delta, \label{tJ-Eb_all}
\end{eqnarray}
where $\Delta=1$ when $\xi<0$ and $\xi'\leq1/2$, and $\Delta=3-\frac{1}{\frac{1}{4}-\zeta^2}$ when $\xi<0$ and $1/2<\xi'<1$.

\section{Conclusions}
\label{Conclusions} \setcounter{equation}{0}

In this paper, we have studied the thermodynamic limit of the one-dimensional supersymmetric $t-J$ with unparallel boundary fields.
It is shown that the contribution of the inhomogeneous term to the ground state energy is inversely proportional with $L$, i.e., $E_{inh}\propto L^{-1}$.
This fact enables us to calculate the surface energy (\ref{TL_Eb}) and (\ref{tJ-Eb_zeta_BDS}),  which is same as that for the case of parallel boundary fields \cite{J.Phys.A:Math.Gen.29.6183}.
Moreover, it implies that the inhomogeneous term in (\ref{T_Q_relation}) surely gives some contributions to the other physical qualities such as the boundary conformal charge,  which are
related to the coefficients in the expansion of energy $E$ in terms of the powers of $L^{-1}$ (namely, the coefficient of $L^{-1}$ corresponds to the conformal charge \cite{Phys.Rev.B54.8491}).

The method used in this paper can be generalized to study
the thermodynamic limit and surface energy of other models related to rational $R$-matrices, such as
the spin-$s$ XXX chain or the $su(n)$ spin chain with unparallel boundary fields. These results may be applied  to the theory of ultra-cold atom systems, asymmetric simple exclusion process.


\section*{Acknowledgments}

We would like to thank Prof. Y. Wang for his valuable discussions and continuous encouragements.
The financial supports from the National Program
for Basic Research of MOST (Grant No. 2016YFA0300600 and
2016YFA0302104), the National Natural Science Foundation of China
(Grant Nos. 11434013, 11425522, 11547045, 11774397, 11775178 and 11775177), the Major Basic Research Program of Natural Science of Shaanxi Province
(Grant Nos. 2017KCT-12 and 2017ZDJC-32) and the Strategic Priority Research Program of the Chinese
Academy of Sciences are gratefully acknowledged. F. Wen also acknowledges the support of the NWU
graduate student innovation fund (No. YYB17003).

\end{document}